\documentclass[twocolumn,showpacs,prb,amsmath,amssymb]{revtex4}
\usepackage{graphicx}
\usepackage{bm}
 
\DeclareMathOperator{\Max}{Max}
\DeclareMathOperator{\Min}{Min}
\begin{document}
\title{Silicon quantum computation based on magnetic
  dipolar coupling} 
\author{Rogerio \surname{de Sousa}}
\affiliation{Department of Chemistry and Pitzer Center for Theoretical
  Chemistry,\\ University of California, Berkeley, CA 94720-1460}
\author{J. D. \surname{Delgado}} \author{S. \surname{Das Sarma}}
\affiliation{ Condensed Matter Theory Center, Department of Physics,
  University of Maryland, College Park, MD 20742-4111} \date{\today}
\begin{abstract}
A dipolar gate alternative to the exchange gate based Kane quantum
computer is proposed where the qubits are electron spins of shallow
group V donors in silicon. Residual exchange coupling is treated as
gate error amenable to quantum error correction, removing the
stringent requirements on donor positioning characteristic of all
silicon exchange-based implementations [B. Koiller {\it et al.}, Phys.
Rev. Lett. {\bf 88}, 027903 (2002)]. Contrary to common speculation,
such a scheme is scalable with no overhead in gating time even though
it is based on long-range dipolar inter-qubit coupling.
\end{abstract}
\maketitle

\section{Introduction}

Since the seminal exchange gate proposal of Loss and DiVincenzo
research on semiconductor spin quantum computation has focused on
implementations based on the electron exchange interaction
\cite{loss98}.  For silicon donor impurities the use of exchange
coupling is problematic since the exchange energy depends sensitively
on donor positioning due to the quantum interference arising from the
six-fold degeneracy of Si conduction band \cite{koiller02}. This
results in the necessity of donor positioning within one silicon bond
($2.4$~\AA) otherwise severe tuning requirements will adversely affect
the scalability of this implementation (in addition to many donor
pairs having nearly zero exchange). This problem is attracting
considerable attention \cite{skinner03} since Si spin quantum computer
architecture is an active research area, and donor spins in
nuclear-spin-free silicon (pure $^{28}$Si) are expected to have very
long coherence times \cite{tyryshkin03,desousa03,abe04} ($T_2\sim T_1 \sim
10^3$~s \cite{feher59}). Here we propose the magnetic dipolar
interaction rather than the exchange interaction between well
separated donor electron spins as a solution to this problem. The
residual exchange interaction is treated as a source of imperfection
in the dipolar gate, whose error probability can be kept below
$10^{-4}$ per operation. Hence the exchange interaction can be ignored
as long as error correction is applied, leading to no necessity of
gate tuning. This is possible due to the long range character of
dipolar coupling (proportional to $1/d^3$, with $d$ the inter-qubit
separation) as opposed to the short-range nature of exchange
[$J\propto d^{2.5}\exp{(-d)}$] \cite{herring64}. Nevertheless this
long-range character led to speculations that a dipolar quantum
computer is not scalable \cite{tahan02}.  We point out that this is
not true, because only up to the 4th nearest neighbor (n.n.)
couplings need to be considered, and highly efficient quantum gates
can be constructed using the method of Leung {\it et al.}
\cite{leung00}, which we develop further to avoid correlated error
between any two qubits inside the error correction manifold.  Similar
pulsing sequences should be useful for a wide variety of solid state
quantum computing architectures based on long range interactions
\cite{scarola03}. The resulting architecture takes advantage of
$g$-factor manipulation and measurement at the single spin level to
avoid the scalability problems inherent to the ensemble NMR proposals
(such as decreasing signal to noise ratio and overlapping resonances
\cite{ladd02,gershenfeld97}).

\section{Dipolar gate fidelity in the presence of 
exchange interaction}

The truncated magnetic coupling between two localized spins is given
by
\begin{equation}
{\cal H}_{12}= \omega_1 S_{1z} + \omega_2 S_{2z} -
\left[D_{12}(\theta,d)-J(a^*,d)\right] S_{1z} S_{2z}. \label{h12}
\end{equation}
Here ${\bm S}_{i}$ are spin-$1/2$ operators, which couple to external
magnetic fields through the Zeeman frequencies $\omega_i=\gamma_i
B_i$. Eq.~(\ref{h12}) is valid provided we neglect terms proportional
to $S_{i\pm}$, which amount to a correction quadratic in
$\left[D_{12}/(\omega_1-\omega_2)\right]$ \cite{mr_books}. Hence a
strong inhomogeneous field is needed (or inhomogeneous gyromagnetic
ratios $\gamma_i$), but since $D_{12}/\gamma_i\lesssim 0.01$~G (see below)
field differences on the $10-100$~G range are sufficient. 
The dipolar interaction is given by 
\begin{equation}
D_{12}(\theta,d)=\frac{\gamma_1
\gamma_2\hbar}{d^3}\left(3\cos^2{\theta}-1\right),
\label{d12}
\end{equation}
where $d$ is the inter-qubit distance and $\theta$ is the angle
between the external magnetic field and the line joining the spins.
The optimum dipolar architecture assumes $\theta=0$, e.g., an array of
spins directed along ${\bf B}$ (this optimal coupling is denoted
$D_{12}$ below). Eq.~(\ref{d12}) has a striking property: If
$\left|\cos{\theta}\right|=1/\sqrt{3}$, the interaction is exactly
zero. Hence in an array of spin qubits dipolar coupling can be
completely suppressed as long as $\pm{\bf B}$ makes one of the
``magic'' angles with the array: $\theta = 54.74^{\circ}$,
$125.26^{\circ}$. Exchange based proposals \cite{loss98} usually
require the donors to be pushed outside the array to switch on the
exchange interaction. In addition, 2D and 3D arrangements have been
considered, particularly to optimize quantum error correction
\cite{divincenzo00,burkard99b}.  In this case it may be impossible to
find a geometry where all bonds are making a magic angle with the $B$
field. Then if dipolar interaction is to be ignored, we will loose
track of the spin state within $10$~$\mu$s (this time should be
considered short if compared with other mechanisms such as nuclear
spectral diffusion \cite{desousa03,abe04}).  Hence dipolar coupling
may have to be taken into account even in exchange gate quantum
computing architectures.  For the exchange interaction between two
hydrogenic donors we use the asymptotic expression \cite{herring64}
\begin{equation}
J(a^*,d)\approx \frac{1.6}{\hbar \varepsilon}\frac{e^2}{a^*}
\left(\frac{d}{a^*}\right)^{5/2}\exp{\left(-2\frac{d}{a^*}\right)},
\label{jherring}
\end{equation}
valid for inter-donor distances $d$ much larger than the Bohr radius
$a^*$. Eq.~(\ref{jherring}) is to be regarded as an envelope for the
strong oscillations of the exchange energy stemming from conduction
band degeneracy \cite{koiller02}. The Bohr radius $a^*$ is related to
the experimental donor ground state energy $E_d$, see
Table~\ref{tab1_spinspin} \cite{note1}.

We will now show how a silicon donor quantum computer can be
implemented with the dipolar interaction and single spin rotations.
The effect of the exchange interaction will be treated as an error,
leading to a lower bound on qubit separation. Assuming $J=0$ in
Eq.~(\ref{h12}), a controlled-Z (CZ) gate is obtained by free
evolution during the time interval
$\tau_{\rm{CZ}}=\pi/D_{12}$ together with $g$-factor shifts \cite{note2},
\begin{equation}
U_{\rm{CZ}}=
\textrm{e}^{-\frac{3\pi}{4}i}
\textrm{e}^{\frac{3\pi}{2}iS_{1z}}
\textrm{e}^{-\frac{\pi}{2}iS_{2z}}
\exp{\left(-i\frac{\pi}{D_{12}}{\cal
H}_{12}\right)}. \label{ucz}
\end{equation}
Below we show how to correct for the Zeeman frequencies $\omega_i$.

We now search for the minimum inter-qubit distance $d$ so that $J$ can
be ignored. A residual exchange interaction $J$ will add an additional
evolution operator to Eq.~(\ref{ucz})
\begin{equation}
U(\alpha)= \exp{\left(-i\alpha S_{1z}S_{2z}\right)},
\end{equation}
with $\alpha=\pi J/D_{12}$. This causes phase error in the CZ gate,
which is better evaluated by looking at two input states orthogonal to
each other. Equivalently we look at the CNOT gate (obtained by a basis
change on the CZ,
$U_{\rm{CNOT}}=e^{-i\frac{\pi}{2}S_{2y}}U_{\rm{CZ}}e^{i\frac{\pi}{2}S_{2y}}$).
Therefore the ``erroneous'' evolution is given by
$\tilde{U}(\alpha)U_{\rm{CNOT}}$, where
\begin{equation}
\tilde{U}(\alpha)=e^{-i\frac{\pi}{2}S_{2y}}U(\alpha)e^{i\frac{\pi}{2}S_{2y}}
\end{equation}
is a $4\times 4$ matrix with elements equal to
$\cos{\left(\alpha/4\right)}$,  $\sin{\left(\alpha/4\right)}$, and
$0$. The error due to a finite $\alpha$ can be evaluated by
calculating the fidelity functions
\begin{equation}
F\{|\psi\rangle,\alpha\}=\left|\langle\psi|U_{\rm{CNOT}}
^{\dag}\tilde{U}(\alpha)U_{\rm{CNOT}}|\psi\rangle\right|,
\label{fid}
\end{equation}
which are simply given by $|\cos{\frac{\alpha}{4}}|$, leading to  an
error probability of $\alpha^2/16$ for small $\alpha$ (the error is
given by $E=1-F^2$). If one wants to ignore exchange interaction, all
that needs to be done is to keep $\alpha^2/16$ less than some critical
bound $p$, for example the 7-qubit encoding threshold
$p=10^{-4}$ \cite{steane96}.  Hence we have $J/D_{12}\leq \sqrt{p}$, or
\begin{equation}
\frac{J}{D}\approx \left(\frac{a^*}{0.02\textrm{\AA}}\right)^2
\left(\frac{d}{a^*}\right)^{11/2}\exp{\left(-2\frac{d}{a^*}\right)}\leq
10^{-2},\label{joverd}
\end{equation}
for $p=10^{-4}$. The length scale for the prefactor in this
expression is given by $\sqrt{2\varepsilon/1.6}\hbar\gamma/e\approx
0.02$~\AA. The range where this inequality is satisfied is
approximately given by $0<d\lesssim 0.03 a^*$ and $d\gtrsim 17
a^*$. The first condition arises due to the divergence of the dipolar
interaction, and is not useful here [also
Eq.~(\ref{jherring}) is only appropriate for $d\gg a^*$]. The physical
solution is the second one, which is optimal (fastest gate) for
$d_{\rm{opt}}\approx 17 a^*$. Table~\ref{tab1_spinspin} shows
$d_{\rm{opt}}$ for various
donors together with their CZ gate time [$\pi/D_{12}$, see
Eq.~(\ref{ucz})].

\begin{table}
\begin{center}
\begin{tabular}{cc c c c c c}
\hline\hline Donor && $E_d$ [meV] & $a^*$ [\AA] & $d_{\rm{opt}}$ [\AA]
& $d_{0}$ [\AA] & $\tau_{\rm{CZ}}$ [$\mu$s] \\\hline Sb && 43 & 18.6 &
315 & 263 & 150\\ P  && 45 & 18.2 & 307 & 256 & 140\\ As && 54 & 16.6
& 279 & 232 & 105\\ Bi && 71 & 14.5 & 241 & 200 & 68\\ \hline\hline
\end{tabular}
\caption{A group V donor electron spin quantum computer, where
  free evolution of the spin-spin dipolar interaction implements CZ
  gates.  Here we show donor electron ground state energies $E_d$
  \cite{lancaster66}, Bohr radius $a^*$ \cite{note1}, optimum
  inter-qubit distance $d_{\rm{opt}}$ (for the exchange interaction to
  be ignored within $10^{-4}$ error probability), inter-qubit distance
  $d_0$ (such that $D_{12}=J$), and the CZ gate times. Fastest gate
  times are obtained for bismuth donors.
\label{tab1_spinspin}}
\end{center}
\end{table}

\section{A scheme for decoupling long-range interactions}

Up to now we have shown that dipolar coupling between two donors can
generate precise two-qubit evolution i.e. a dipolar coupled-qubit Si
gate can be constructed. However the situation becomes complicated
when we consider an array of many donors. Particularly the long range
nature of the dipolar interaction implies every spin in the array will
be coupled to each other, raising questions about the scalability of
this proposal (This was one of the original motivations for
introducing the exchange gate since exchange can be exponentially
suppressed by electrically controlling wave function overlap). For
example, it is possible that the complexity of the pulsing sequences
(leading to the desired quantum algorithm) might scale exponentially
with the number of qubits, effectively making the problem of
determining the evolution as hard as any mathematical problem a
quantum algorithm is constructed to solve \cite{tahan02}. Nevertheless
this is not true for the case considered here, because using the same
argument leading to the discard of the exchange interaction we can
neglect (within the $10^{-4}$ threshold) dipolar coupling between any
spin and its 5th or higher n.n. [By Eq.~(\ref{joverd}),
$D_{1k}/D_{12}=1/k^3$, which is less than $10^{-2}$ for $k\geq
5$]. Hence Eq.~(\ref{h12}) generalized to a 1D spin array is
\begin{equation}
{\cal H}=\sum_i \omega_i S_{iz} - \frac{1}{2}\sum_{i;j=i-4}^{i+4} D_{ij}
S_{iz}S_{jz},\label{hij}
\end{equation}
where $i$ is an integer labeling the location of each donor ($i$ is
assumed positive as well as negative). The finite coupling range allows us to
develop quantum gates using a sequence of $\pi$-pulses applied to
subsets of the spins [each $\pi$-pulse is given by
$X^{(i)}=\exp{(i\pi S_{ix})}$]. The key point is that
the interaction between any two spins can be canceled 
using two $\pi$-pulses \cite{mr_books},
\begin{equation}
\exp{\left(-i \frac{\tau}{2}
  DS_{1z}S_{2z}\right)}X^{(2)}\exp{\left(-i \frac{\tau}{2}
  DS_{1z}S_{2z}\right)} X^{(2)}=I,\label{ref12}
\end{equation}
where $I$ is the identity operator. Our task is now to find the
``decoupling'' scheme which completely refocuses Eq.~(\ref{hij}) after
some time interval $\tau$ (therefore enabling single qubit rotation on
any spin) and also to produce sequences for ``selective recoupling'',
which provide CZ evolution for any n.n. pair. For this task we use the
method of Ref.~\onlinecite{leung00} which consists in constructing
sign matrices $S_n$ representing the $\pi$-pulses. A $n\times m$ sign
matrix has each element equal to $\pm 1$ (denoted simply by $\pm$),
and correspond to a system of $n$ spins where evolution during a time
$\tau$ is divided into $m$ time intervals. If spin $i$ has its
interaction reversed in any $l$-th time interval [by application of
$X^{(i)}$ before and after this time interval], then
$(S_{n})_{il}=-1$, otherwise $(S_{n})_{il}=+1$.  For example,
Eq.~(\ref{ref12}) corresponds to
\begin{equation}
S_2=
\left(
\begin{array}{cc}
+&+\\
+&-
\end{array}
\right).\label{h2}
\end{equation}
The interaction between two spins $i,j$ is decoupled if the rows $i$
and $j$ of $S_n$ disagree in sign for half of the $m$ time intervals.
Equivalently, the inner product between these rows is zero. This
property leads to a connection with the theory of Hadamard matrices:
$H_{\overline{n}}$ is a $\overline{n}\times \overline{n}$ Hadamard
matrix if and only if $H_{\overline{n}}\cdot
(H_{\overline{n}})^{T}=\overline{n}I$. Hence a possible solution for
the decoupling problem of $n$ spins is to construct $S_{n}$ from $n$
rows out of a Hadamard $H_{\overline{n}}$ where $\overline{n}\geq n$.
Actually such a solution turns out to be the most efficient one (the
smallest $\overline{n}$ satisfying $\overline{n}\geq n$ is the minimum
number of intervals $m$ in the set of possible $S_n$) because one can
show that it is impossible to add an additional row orthogonal to
$H_{\overline{n}}$. Hadamard matrices exist for $\overline{n}=1$, $2$
[Eq.~(\ref{h2})], $4$, $8$, $12$,~$\ldots$ (see
Ref.~\onlinecite{leung00} and references therein). The finite coupling
range of Eq.~(\ref{hij}) suggests $\overline{n}=12$ as a convenient
solution.  $S_n$ can be assembled as a $n\times 12$ matrix composed of
7 ordered rows from $H_{12}$ (identical rows are 7 rows apart).
For the particular case of $n=14$ (general $n$ is obtained by row
repetition),
\begin{equation}
S_{14}=
\left(
\begin{array}{cccccccccccc}
+&+&+&+&+&+&-&-&-&-&-&-\\
+&+&+&-&-&-&+&+&+&-&-&-\\
+&-&-&+&+&-&-&+&+&-&-&+\\
+&+&-&+&-&-&+&-&-&+&-&+\\
+&+&-&-&+&-&-&+&-&+&+&-\\
+&+&-&-&-&+&-&-&+&-&+&+\\
+&-&+&+&-&-&-&-&+&+&+&-\\
\hline
+&+&+&+&+&+&-&-&-&-&-&-\\
+&+&+&-&-&-&+&+&+&-&-&-\\
+&-&-&+&+&-&-&+&+&-&-&+\\
+&+&-&+&-&-&+&-&-&+&-&+\\
+&+&-&-&+&-&-&+&-&+&+&-\\
+&+&-&-&-&+&-&-&+&-&+&+\\
+&-&+&+&-&-&-&-&+&+&+&-\\
\end{array}
\right).\label{s14}
\end{equation}
Here we extract the first (and last) 7 rows of $S_{14}$ from $H_{12}$ with
the first row ($++\ldots$) excluded so that Zeeman splitting is also
canceled. $S_{14}$ requires a total of 80 $\pi$-pulses which
are applied in $12$ sets (less than 14 pulses are applied in each set
because $X^{(i)2}=I$ -- hence no rotations need to be applied when the
sign is the same for neighboring time intervals). An array of $n$
spins will require less than $6 n$ pulses. Selective recoupling is
achieved by choosing identical rows for the spins which are to be
coupled. These rows are chosen from the 4 remaining rows of $H_{12}$, 
for example
\begin{equation}
S'_{14}=
\left(
\begin{array}{cccccccccccc}
+&+&+&+&+&+&-&-&-&-&-&-\\
+&+&+&-&-&-&+&+&+&-&-&-\\
+&-&-&+&+&-&-&+&+&-&-&+\\
+&+&-&+&-&-&+&-&-&+&-&+\\
+&+&-&-&+&-&-&+&-&+&+&-\\
\bm{+&-&+&-&+&-&+&-&-&-&+&+}\\
\bm{+&-&+&-&+&-&+&-&-&-&+&+}\\
\hline
+&+&+&+&+&+&-&-&-&-&-&-\\
+&+&+&-&-&-&+&+&+&-&-&-\\
\bm{+&-&+&-&-&+&-&+&-&+&-&+}\\
\bm{+&-&+&-&-&+&-&+&-&+&-&+}\\
+&+&-&-&+&-&-&+&-&+&+&-\\
+&+&-&-&-&+&-&-&+&-&+&+\\
+&-&+&+&-&-&-&-&+&+&+&-\\
\end{array}
\right)\label{sr14}
\end{equation}
implements CZ operations between spins 6, 7 and 10, 11 in parallel
(bold).  We point out that each 7-qubit structure in Eqs.~(\ref{s14})
and (\ref{sr14}) form an error correction block for the Steane code
\cite{steane96}. Note that residual dipolar interaction couples qubits
in different blocks. This is important for the assumption of
uncorrelated errors within each block and the validity of the
$10^{-4}$ threshold (see section IV). The spurious couplings lead to
error of the order of $7^{-6}\sim 10^{-5}$ in Eq.~(\ref{s14}) and
$22^{-6}\sim 10^{-8}$ for selective recoupling when all blocks execute
CZ in parallel (massive parallelization of CZ is needed for efficient
computation and quantum error correction).

Therefore the complete gate time for a large 1D array is the same as
for two donors (approximately $100$~$\mu$s, being optimal for bismuth
-- see Table~\ref{tab1_spinspin}). This shows that a dipolar donor
electron spin quantum computer is reliable: If the silicon lattice is
isotopically purified (free of $^{29}$Si nuclear spins), the coherence
time will be limited by the spurious exchange and dipolar couplings,
with a quality factor of the order of $10^4$.  A key advantage of this
architecture is the inter-qubit distance, which is three times larger
than other proposals for donors \cite{loss98}. Also there is no need
for an inter-qubit ``J'' gate, or any electrical control over wave
function overlap \cite{loss98,skinner03}. This should make gate
lithography much simpler (one needs to incorporate $g$ factor control
\cite{salis01} and single spin measurement/initialization
electrodes \cite{kane00b} on top of each donor).

The considerations above can be generalized to any long-range coupling
$D\propto 1/d^r$. The number of n.n.  which need to be decoupled is
given by $\Max{(k)}\leq p^{-1/2r}$, where $p$ is the desired error
probability. Hence $\tau_{\rm{CZ}}$ needs to be broken into
$\overline{n}\sim p^{-1/2r}$ time intervals.  For example, $r\gtrsim
1$ and $p=10^{-4}$ leads to $\overline{n}\sim 100$. Implementation of
any quantum gate is possible as long as the time for single spin
rotation is much less than $\tau_{\rm{CZ}}/\overline{n}$. The dipolar
case considered here clearly satisfies this criterion, since
$\tau_{\rm{CZ}}/\overline{n}\sim 10$~$\mu$s (rotation times of the
order of $0.1$~$\mu$s are easily achievable). Finally, notice that
this approach for decoupling can also be applied to general
anisotropic exchange interactions, since these can be transformed into
the $S_{iz}S_{jz}$ form by appropriate spin rotations.

\section{Error correction of residual long-range coupling}

Here we show how imperfections arising from spurious long range
couplings connecting qubits in \emph{distinct quantum error correction
  blocks} can be corrected by the usual syndrome diagnosis (projective
measurement on each block)\cite{nielsen00}.  The proof presented here
is based on the simplest error correction code, the ``three bit flip
code'' (section 10.1.1 of Ref.~\onlinecite{nielsen00}). However, we
emphasize that these results are easily extended to the complete 7-bit
Steane's code \cite{steane96} which correct for any type of continuous
error on each qubit within its block. The essence of our proof is that
the syndrome measurements on each block effectively destroys error
correlation between qubits belonging to different blocks.

Consider two error
correction blocks constituted by qubits 1,2,3 (first block) and 4,5,6
(second block). The residual coupling Hamiltonian is
\begin{equation}
{\cal H}=-4c' \left(
S_{1z}S_{4z}+S_{2z}S_{5z}+S_{3z}S_{6z}
\right),\label{h}
\end{equation}
and the evolution operator after one ``clock time" $\tau$ is
\begin{eqnarray}
{\cal U}(\tau)&=&\exp{\left(-i{\cal H}\tau\right)}
=\cos^3{(c)}I + i \sin{(c)}\cos^2{(c)}\nonumber\\
&&\times\left[
\sigma_{1z}\sigma_{4z}+
\sigma_{2z}\sigma_{5z}+
\sigma_{3z}\sigma_{6z}
\right]-\cos{(c)}\sin^2{(c)}\nonumber\\
&&\times\left[
\sigma_{1z}\sigma_{2z}\sigma_{4z}\sigma_{5z}+
\sigma_{1z}\sigma_{3z}\sigma_{4z}\sigma_{6z}+
\sigma_{2z}\sigma_{3z}\sigma_{5z}\sigma_{6z}
\right]\nonumber\\
&&-i\sin^3{(c)}\left[
\sigma_{1z}\sigma_{2z}\sigma_{3z}
\sigma_{4z}\sigma_{5z}\sigma_{6z}
\right].
\label{utau}
\end{eqnarray}
Here $c=c'\tau=\pi D_{14}/D_{12}$ is much less than one (in the case
of 7-qubit blocks $c\sim 7^{-3}$). To map this problem into the
bit-flip code we use the y basis for our spin qubits:
\begin{eqnarray}
|0\rangle &=& |+y\rangle =\frac{1}{\sqrt{2}}\left(\mid\uparrow\rangle 
+i\mid\downarrow\rangle\right),\\
|1\rangle &=& |-y\rangle =\frac{1}{\sqrt{2}}\left(\mid\uparrow\rangle 
-i\mid\downarrow\rangle\right).
\end{eqnarray}
In this basis the Pauli matrices of Eq.~(\ref{utau}) act as a bit flip
operator ($\sigma_z |0\rangle =| 1\rangle$, $\sigma_z | 1\rangle =|
0\rangle$). Eq.~(\ref{utau}) contains three contributions: (1)
simultaneous bit flips of one spin in block 1 and another in block 2.
This leads to error probability of $\sim c^2$ for each spin in each
block [square of the amplitude, see Eq.~(\ref{fid})].  Note that error
is correlated between blocks; (2) Simultaneous double bit flip in both
blocks, with probability equal to the square of the single bit flip
probability ($\sim c^4$). This process has the same order of magnitude
of two independent single bit flip errors occurring at the same time.
This type of error is only corrected after two concatenations of error
correction are in place; (3) three bit flip error in both blocks,
probability is cube of single bit flip, equivalent to three
simultaneous independent bit flips.

We start by assuming blocks 1 and 2 store the state
\begin{equation}
|\psi(0)\rangle=a_1 |00\rangle + a_2 |01\rangle +a_3 |10\rangle +a_4 |11\rangle.
\end{equation}
For fault tolerant quantum computing, we add 4 additional ancilla
qubits which encode the state as
\begin{eqnarray}
|\psi(0)\rangle&=&a_1 |000,000\rangle + a_2 |000,111\rangle\nonumber\\
&&+a_3 |111,000\rangle +a_4 |111,111\rangle.
\end{eqnarray}
Time evolution under the spurious coupling Hamiltonian
[Eq.~(\ref{utau})] yields
\begin{eqnarray} 
|\Psi(\tau)\rangle &=& \cos^3{(c)} |\Psi(0)\rangle -i\sin^3{(c)}
 |\phi\rangle+i\sin{(c)}\cos^2{(c)}\nonumber\\
&&\times\left\{
a_1\left[ |100,100\rangle + i\tan{(c)}|011,011\rangle\right]\right.\nonumber\\
&&+a_2\left[ |100,011\rangle + i\tan{(c)}|011,100\rangle\right]\nonumber\\
&&+a_3\left[ |011,100\rangle +
i\tan{(c)}|100,011\rangle\right]\nonumber\\
&&\left.+a_4\left[ |011,011\rangle +
i\tan{(c)}|100,100\rangle\right]\right\}\nonumber\\
&&+\cdots,
\label{psitau}
\end{eqnarray}
where the swapped state $|\phi\rangle$ is given by
\begin{eqnarray}
|\phi\rangle&=&a_1 |111,111\rangle + a_2 |111,000\rangle\nonumber\\
&&+a_3 |000,111\rangle +a_4 |000,000\rangle.
\end{eqnarray}
Error correction proceeds with projection measurements over the
syndromes 0,1,2, and 3 in each block $k=1,2$:
\begin{eqnarray} 
P_0^{(k)} &=& |000\rangle\langle 000| + |111\rangle\langle 111|,\\
P_1^{(k)} &=& |100\rangle\langle 100| + |011\rangle\langle 011|,\\
P_2^{(k)} &=& |010\rangle\langle 010| + |101\rangle\langle 101|,\\
P_3^{(k)} &=& |001\rangle\langle 001| + |110\rangle\langle 110|.
\end{eqnarray}
Depending on the outcome of the measurement we apply the corresponding
correction operator $U_i^{(k)}$ (for example, $U_0^{(1)}=I$,
$U_1^{(1)}=\sigma_{1z}$, $U_2^{(1)}=\sigma_{2z}$,
$U_3^{(1)}=\sigma_{3z}$). The final corrected density matrix is an
incoherent superposition of each possible error:
\begin{eqnarray}
\rho_c = \sum_{i,j} U_{i}^{(1)}P_{i}^{(1)}U_{j}^{(2)}P_{j}^{(2)} 
|\Psi(\tau)\rangle\nonumber\\
\times\langle\Psi(\tau)|
P_{i}^{(1)\dag}U_{i}^{(1)\dag}
P_{j}^{(2)\dag}
U_{j}^{(2)\dag}.
\end{eqnarray}
Because error is correlated between blocks the projection
$P_{i}^{(1)}P_{j}^{(2)}$ onto state $|\Psi(\tau)\rangle$ is zero unless
$i=j$. For example, 
\begin{eqnarray}
\prod_{k=1}^{2} U_{0}^{(k)}P_{0}^{(k)}
|\Psi(\tau)\rangle &=&
\cos^3{(c)}|\Psi(0)\rangle-i\sin^3{(c)}|\phi\rangle,\nonumber\\
\prod_{k=1}^{2}U_{l}^{(k)}P_{l}^{(k)}
|\Psi(\tau)\rangle &=&
i\sin{(c)}\cos^2{(c)}\nonumber\\
&&\times\left[|\Psi(0)\rangle+i\tan{(c)}|\phi\rangle\right],
\end{eqnarray}
for $l=1, 2, 3$ [Note that Eq.~(\ref{psitau}) omitted the syndrome
subspaces 2, 3].
Finally it is straightforward to calculate the fidelity squared,
\begin{eqnarray} 
|F|^2&=&\langle\Psi(0)|\rho_c||\Psi(0)\rangle\nonumber\\
&&=\cos^4{(c)}\left[1+2\sin^2{(c)}\right]\nonumber\\
&&+\sin^4{(c)}\left[1+2\cos^2{(c)}
\right]|\langle\Psi(0)|\phi\rangle|^2.
\label{fidpsi}
\end{eqnarray}
Maximum error occurs when the second term of Eq.~(\ref{fidpsi}) is
zero. This leads to 
\begin{equation}
\Max{(E)}=1-\Min{(|F|^2)}\approx 3c^4 +{\cal O}(c^6),
\end{equation}
which is the square of the error without error correction.  Hence
neglecting small dipolar coupling between different error correction
blocks is for all practical purposes equivalent to having a source of
independent uncorrelated error identical to the one assumed in the
quantum error correction literature.

\section{Discussion}

We now consider the feasibility of our dipolar QC proposal for III-V
semiconductor donor impurities and quantum dots.  Although these
materials have a small effective mass (implying higher $a^{*}$ and
$d_{\rm{opt}}$), some of the narrow gap semiconductors have quite
large bulk $g$ factors, enhancing dipolar coupling. A simple
estimation is obtained from the relation $\tau_{\rm{CZ}} \sim
(0.3/m^{*})^{3} (2/g)^{2} \times 100~\mu\rm{s}$. Using the parameters
of Ref.~\onlinecite{desousa03b} we get $\tau_{\rm{CZ}} \sim 0.1$~s for
GaAs and $\tau_{\rm{CZ}} \sim 1$~ms for GaSb, InAs, and InSb donors
impurities (quantum dots have dipolar gate times higher by
approximately a factor of 10 due to larger Bohr radii). Hence our
proposal is not feasible for GaAs, but might work for the narrow gap
III-V materials as long as decoherence due to nuclear spectral
diffusion is suppressed by nuclear polarization \cite{desousa03}. In
this case spin-flip followed by phonon emission will be the dominant
decohering process. Adjusting the external magnetic field, coherence
times of the order of a few seconds are achievable \cite{desousa03b},
suggesting the possibility of quality factors greater than $10^3$
in a narrow gap donor dipolar quantum computer, which does not require
exchange interaction control and can be constructed with current
lithography techniques.

In conclusion we consider a quantum computer architecture based on
dipolar-coupled donors in silicon. Although gate times are
considerably longer than exchange-based implementations (albeit same
time scales as the solid state NMR proposals \cite{ladd02}), one does
not need atomic precision donor implantation or electrical control of
two-qubit couplings. Particularly ``top-down'' construction schemes
based on ion implantation should benefit from our proposal, because
these lack precision in donor positioning in addition to creating
interstitial defects \cite{schenkel03} (dipolar coupling is nearly
insensitive to electronic structure).  Our proposal for decoupling of
short-range ``always on'' interactions together with error correction
of the remaining long-range couplings apply equally well to any solid
state implementation based on other types of long-range interactions
(as long as the coupling is bilinear) \cite{ladd02,scarola03}, opening
the way to implementations which do not have severe lithography
requirements.  We acknowledge useful discussions with B. E. Kane, J.
Kempe, T.D.  Ladd, T. Schenkel, J. Vala and W.  Witzel. This work is
supported by ARDA, LPS, US-ONR, and NSF.

%
%


\begin{thebibliography}{99}
\bibitem{note1} This is done by introducing an effective charge $Z$
for the donor Coulomb interaction.  Relating $Z$ to the experimental
value $E_d$ we get $a^*(E_d)= \sqrt{26\;\rm{meV}/|E_d|}\times
23.9\;\textrm{\AA}.$
\end{thebibliography}
\end{document}